# MiftyCoin (MFT):

# A Cryptocurrency Mined with Proof of Human Work

Javier A. Arroyo-Figueroa

Entevia, LLC [1]

Email: jarroyo@entevia.com

## Abstract

We present in this paper a cryptocurrency called Mobile Fungible Token (MFT) or "MiftyCoin", which is mined with Proof of Human Work (PoH). Blocks in MFT's blockchain are mined by users solving unique 24-tile puzzles autogenerated as a function of block hash values. Each tile in the puzzle is a 4-sided domino-like square, where the number of dots per square is a function of a subset of the bits of a corresponding byte in the block's hash value. The objective is to find a set of tile moves that end up in an arrangement with an optimal score; where each matching tile side increases the score by one. The block with the highest score gets a reward.

More information about MiftyCoin is available at: https://www.miftycoin.com.

## 1. Introduction

Since Satoshi Nakamoto's seminal work on Bitcoin [N08], cryptocurrencies and blockchain technologies have flourished into numerous ecosystems that go from simple verification networks to complex banking and trading entities. A high number of cryptocurrencies have emerged since, with Bitcoin at a lead with a total market capitalization in the order of hundreds of billions of dollars [T20].

Despite advances in the field, most cryptocurrencies follow Bitcoin's block validation scheme of Proof of Work (PoW). In this scheme, a high number of network nodes (commonly known as "miners") compete by trying to solve a computationally-expensive problem that is a function of the block's hash value- a process called "mining". The computational effort spent in mining is not only time-consuming, but

---

energy-consuming as well. It is estimated that between 120 and 240 billion kilowatt-hours per year are consumed in mining; a range that exceeds the total annual electricity usage of many individual countries, such as Argentina or Australia [W22].

An energy-saving alternative to PoW is Proof of Stake (PoS), which requires miners to bet an amount of the same cryptocurrency to get a reward from a pool; the miner that bets the most is rewarded and the mined block is marked as valid. Although being more energy-efficient than PoW, this approach has the disadvantage that large amounts of cryptocurrency is placed at risk; small miners are discouraged from participating, concentrating the mining power in a few. A more "democratic" or "socialized" mining scheme would be preferable, while keeping in mind the objective of not harming the environment.

An alternative to PoW and PoS that has been proposed in literature is Proof of Human Work (PoH) [BZ16]. With PoH, the mining effort is spent by humans, not machines. Assuming that the only power consumed by human effort are calories and the by-product's carbon emissions are low, PoH can be considered a "green" technology- with the additional advantage of being more democratic/socialized than PoS.

Seminal proposals on PoH were based on the generation of "captchas" where the captcha-generation function is irreversible via obfuscation. Unfortunately, later works proved that such obfuscation is impossible [B12].

Our motivation for a different PoH scheme is one that: (i) does not rely on obfuscation; (ii) produces a unique puzzle per solver, such that solutions cannot be shared; (iii) has the goal of finding an optimal solution, which requires exponential-time complexity when solved by a machine; and (iv) produces solutions that can be verified in polynomial time.

We present in this paper a cryptocurrency called Mobile Fungible Token (MFT) or "MiftyCoin", which is mined by users solving unique 24-tile puzzles autogenerated as a function of block hash values. Each tile in the puzzle is a 4-sided domino-like square, where the number of dots per square is a function of a subset of the bits of a corresponding byte in the block's hash value. The objective is to find a set of tile moves that end up in an arrangement with an optimal score; where each matching tile side increases the score by one. The block with the highest score gets a reward (constrained by a few rules, as it will be presented in this paper).

Works on applications of AI to the 24-tile puzzle have focused on solving a puzzle with a single goal (the target state) and, in some instances, finding an optimal (shortest) solution path (e.g. minimum number of moves). However, none have considered the concept of an optimal state (score). It is known that the state space of the 24-tile puzzle is in the order of $10^{25}$ states. Intuitively, finding an optimal solution to the puzzles generated by MFT is an NP-Complete problem, thus no efficient algorithm has been found; this is where human capacity surpasses machines.

The rest of this paper is organized as follows. In the next section, we present the system architecture. The consensus mechanism is presented in Section 3, followed by an explanation of the Proof of Human Work (PoH) in Section 4. Puzzle generation is discussed in Section 5. Our conclusions are presented in Section 6.

## 2. Architecture

The architecture of the MFT network is presented in Figure 1. It is comprised of three layers: (i) the Client Layer, consisting of mobile phones connected to the network through trusted nodes (or T-nodes); (ii) the Trusted Layer, comprised of multiple T-nodes; and (iii) the Blockchain (trustless) Layer, comprised of blockchain nodes (or B-nodes).

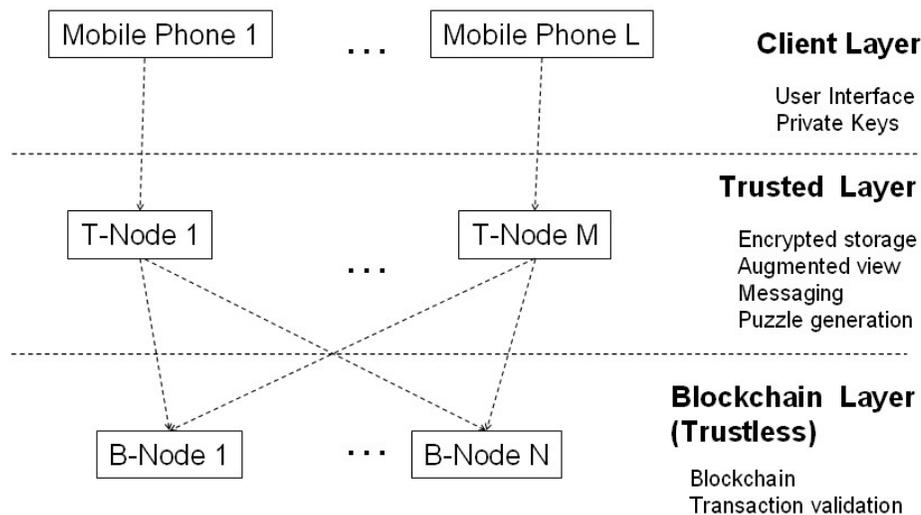

**Figure 1.** The 3-layered architecture of MFT

Like in any blockchain network, nodes in the Blockchain Layer, or B-nodes, are responsible for maintaining a consistent global ledger in a blockchain; this is also referred to as the trustless layer, as it is designed to handle communications in a decentralized network where no node can be trusted and data consistency is achieved by a consensus mechanism.

There are two economic incentives to maintain B-nodes: (i) they compete with humans and can get rewarded by solving puzzles, which also helps in maintaining the blockchain updated when no humans are available to solve puzzles; and (ii) they also act as *brokers* and charge a fee when accepting transactions, the amount of which is determined by market dynamics.

On top of the Blockchain Layer lies the Trusted Layer which, as its name implies, contains nodes that can be trusted and handle encrypted key storage, augmented view of transactions (e.g. transaction descriptions which are not supported in the blockchain), messaging and puzzle generation.

Mobile devices do not interact directly with the Blockchain Layer, but rather with the Trusted Layer, as the later provides functions needed by applications that are missing from the former. For security, private keys are stored in the Client Layer.

## 3. Consensus Mechanism

Since the MFT blockchain is maintained by a network of trustless nodes, a consensus mechanism is used. A sketch of this mechanism is presented here. Each node should follow these consensus rules:

1. The node's clock must be synchronized using NTP [I10].
2. A new block is mined every 10 minutes, specifically at the hour and at 10,20,30,40 and 50 minutes after each hour; and regardless of whether there are pending transactions or not.
3. A block must contain: (i) a unique block identifier; (ii) a block number; (iii) the hash of the previous block in the blockchain (or zero if it is the first block in the chain); (iv) a list of identifiers of transactions included in the block; (v) the address of the block's puzzle solver; (vi) the block's "magic number", which is the minimum integer value that produces a block hash that brings a zero-score puzzle; (vii) a timestamp; (viii) the block's hash value; (ix) the starting puzzle, represented as a sequence of bits; (x) the puzzle's solution, represented as sequence of steps; (xi) the ending puzzle, which is the puzzle after applying the solution; and (xii) the puzzle score.

4. A block must have: (i) a valid hash value; (ii) a valid starting puzzle, which is a zero-score puzzle generated from the block's hash value; (iii) a valid magic number, which is the minimum integer value that produces a zero-score puzzle; (iv) a valid solution, which is list of valid tile moves; (v) a valid ending puzzle, which is the puzzle after applying the moves to the starting puzzle.

5. The modulo-10 of each block's timestamp minutes value in UTC (number of milliseconds since the start of epoch, divided by 60,000) should be equal to zero; in other words, it should be a multiple of 10.

6. The timestamp of a block should be greater than the timestamp of its predecessor.

7. At the time a block-mining request is received by a node: (i) the block's timestamp value cannot be more than 30 seconds ahead of the node's timestamp; (ii) the block's timestamp cannot be five minutes older than the node's timestamp.

8. The timestamp of transactions in a block cannot be older than 40 minutes from the block's timestamp, and cannot be closer than eight minutes from the node's timestamp.

9. A block is marked as *optimal* (among all blocks with the same number) if: (i) it has the highest chain length, which is the total number of transactions in its blockchain; (ii) it has the highest puzzle score; and (iii) it has the lowest hash value.

10. There is only one optimal block within blocks with the same number;

11. The blockchain is formed only with optimal blocks.

12. At the time a transaction-creation request is received by a node: (i) the transaction's timestamp value cannot be more than 30 seconds ahead of the node's timestamp; (ii) the transaction's timestamp cannot be 60 minutes older than the node's timestamp.

13. A reward transaction for a block A can only be accepted if it is included in a block B that is no more than three blocks away from A in the blockchain.

We will present in more detail the Proof of Human Work (PoH) and puzzle generation schemes.

# 4. Proof of Human Work (PoH)

A block is mined by a human by solving a 24-tile puzzle that is automatically generated from the block's hash value. An example of a puzzle is presented in **Error! Reference source not found.**. The score of a puzzle is computed by adding, for each tile, the number of matching sides with adjacent tiles. The objective is to maximize the puzzle score, which is done by moving the tides along. The total number of moves cannot exceed 125. The winner is rewarded with MFT$50 (minus the broker fees).

Notice that there is no single solution to a puzzle. Since the objective is to maximize the puzzle score, it is an optimization problem with a large state space, which is hard to solve by a machine, given the limitations of today's technology.

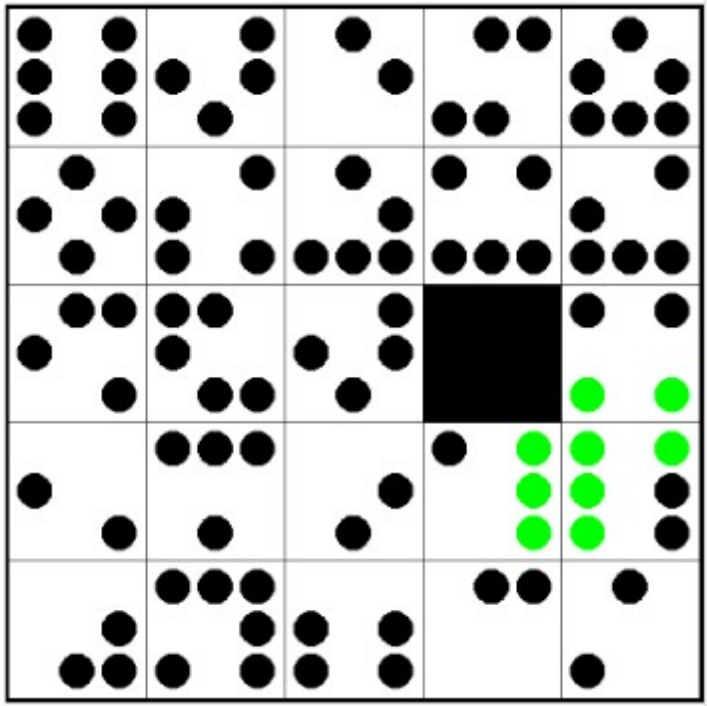

**Figure 2.** An example of a puzzle with score = 4

A *mining competition* occurs every ten minutes, with intervals at 0, 10,20,30,40 and 50 minutes after the hour.

## 5. Puzzle Generation

A puzzle for a block is automatically generated from the block's hash value. The hash value of a block is the SHA-256 hash of the concatenation of the values of the block fields in textual form, where each character is encoded in Unicode. This concatenation is made from the following fields and in the following order:

1. Block number;
2. Timestamp;

3. Magic number;

4. Previous-block hash;

5. Puzzle solver's address;

6. Concatenation of the hash values of the transactions included in the block;

The following is a sketch of the puzzle-generation algorithm:

1. Set the magic number M=0.

2. Number each tile in sequence, starting with number 1 at the top-left corner, always increasing by one, moving to the right in the first row, then moving back to the leftmost tile in the next row and continuing the pattern until tile number 24.

3. Designate eight bits [b0 … b7] in each tile using the following pattern:

   ```
   b0 b1 b2
   b7    b3
   b6 b5 b4
   ```

4. Number each byte in the block hash, starting with number 1 for the leftmost byte (assuming a little endian notation).

5. For each tile in the puzzle:

   a. Let *i* be the tile's number;

   b. Let B be the block hash's *i*th byte.

   c. Let the tile's [b0…b7] correspond to the eight bits of B, where b0 is B's most significant bit.

   d. Each bit that equals one is represented by a dot in the tile; otherwise it is represented by a blank (absent dot).

6. Compute the puzzle score by adding, for each tile, the number of matching sides with adjacent tiles.

7. If the score is not zero, increase M and repeat the process starting in Step #5, until the score is zero.

# 6. Conclusion

We have presented MiftyCoin, a cryptocurrency mined with Proof of Human Work (PoH). Human effort is placed by solving unique 24-tile puzzles autogenerated as a function of block hash values. The incentive is to get a reward of MFT$50. Nodes in the blockchain network have the incentive of getting a reward by competing with humans and by charging broker fees.